\documentclass[american,twocolumn,showpacs,preprintnumbers,amsmath,amssymb,prl]{revtex4-1}
\usepackage[T1]{fontenc}
\usepackage[latin9]{inputenc}
\usepackage{amsmath}
\usepackage{graphicx}
\usepackage{esint}
\usepackage{hyperref}
\usepackage[FIGTOPCAP, footnotesize, raggedright,tight]{subfigure}

\makeatletter
\@ifundefined{textcolor}{}
{%
 \definecolor{BLACK}{gray}{0}
 \definecolor{WHITE}{gray}{1}
 \definecolor{RED}{rgb}{1,0,0}
 \definecolor{GREEN}{rgb}{0,1,0}
 \definecolor{BLUE}{rgb}{0,0,1}
 \definecolor{CYAN}{cmyk}{1,0,0,0}
 \definecolor{MAGENTA}{cmyk}{0,1,0,0}
 \definecolor{YELLOW}{cmyk}{0,0,1,0}
 }

\usepackage{graphicx}
\usepackage{dcolumn}
\usepackage{bm}
\usepackage[usenames,dvipsnames]{color} 

\makeatother

\usepackage{babel}

\newcommand{\upd}{\, \mathrm{d}}
\newcommand{\fbrac}[1]{\!\left(#1\right)}
\newcommand{\slfrac}[2]{\left.#1\middle/#2\right.}

\begin{document}

\title{Stability of localized wave fronts in bistable systems}
\author{Steffen Rulands}
\author{ Ben Kl\"{u}nder }
\author{Erwin Frey}
\email{frey@lmu.de}
\affiliation{Arnold Sommerfeld Center for Theoretical Physics and Center for NanoScience, Department of Physics, Ludwig-Maximilians-Universit\"at M\"unchen, Theresienstr. 37, D-80333 M\"unchen, Germany}
\date{\today}
\begin{abstract}
Localized wave fronts are a fundamental feature of biological systems from cell biology to ecology. Here, we study a broad class of bistable models subject to self-activation, degradation and spatially inhomogeneous activating agents. We determine the conditions under which wave-front localization is possible and analyze the stability thereof with respect to extrinsic perturbations and internal noise. It is found that stability is enhanced upon regulating a positional signal and, surprisingly, also for a low degree of binding cooperativity. We further show a contrasting impact of self-activation to the stability of these two sources of destabilization.
\end{abstract}
\maketitle

Bistable systems are ubiquitous in nature. For example, genetic switches are bistable systems that store the activation state of a gene \cite{Alberts-B:2002fk,PhysRevLett.106.088101}. In population dynamics, a minimum population size is often needed to establish a stable population~\cite{Stephens:1999}.  The spatial versions of such systems admit traveling wave solutions, \emph{e.g.} describing the outbreak of viruses or the colonization of territory~\cite{Lewis1993141,o2000mathematical,ELE:ELE787,Das:2009}. If bistable systems are subject to external spatial gradients, traveling waves  may localize in restricted spatial domains~\cite{Keitt2001,Mori2008,Mori2010}. An important example arises in early embryogenesis of \emph{Drosophila melanogaster} where maternal morphogen gradients provide positional information for  gene regulation \cite{Driever:1988-1,Driever:1988-2,Lopes2008,Tostevin2007}. The morphogen Bicoid is present as a monotonically decreasing concentration in the embryo and controls the step-like activation of the gene Hunchback, which also enhances its own activation effectively producing a bistable system. The exact position of the Hunchback front is pivotal to the embryo's fate \cite{Lopes2008}. Hence, the front's stability to extrinsic perturbations or internal noise is paramount. Wave localization and the stability of the front also play an important role  in other contexts. In ecology, birth rates may have spatial dependence, \emph{e.g.} due to spatial variance in temperature or resource availability~\cite{Putman1984,Czaran1998}. The localized boundaries between species are subject to large fluctuations due to the low number of individuals in the boundary region. This may eventually lead to the extinction of one of the species due to demographic stochasticity \cite{Meerson2011a}. Last, in bio-technological applications, this mechanism might be used to create localized fronts of proteins~\cite{Loose2011}. 

Motivated by these processes, we investigate a broad class of bistable diffusion-reaction models with reaction terms comprising self-activation, external activation, and degradation. While self-activation and degradation are assumed to be spatially uniform, the external activation is taken to be position-dependent. We consider two qualitatively different types of external gradients and determine the parameter range for which wave localization is possible. Moreover, we ask how stable these localized fronts are with respect to extrinsic and intrinsic noise, and we determine optimality conditions minimizing the front's susceptibility to such perturbations. 

\begin{figure*}[ht!]
\begin{center}
\includegraphics[width=0.95\textwidth]{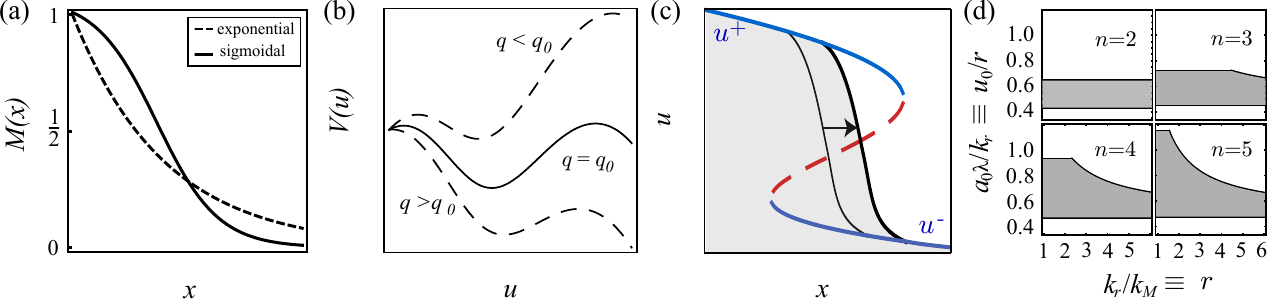}
\caption{\label{fig:box1}(Color online)  (a) Two types of gradients: exponential decay (dashed line) and a sigmoidal profile ensuing from regulating an exponentially decaying input (solid line). (b) The potential for different values of the front position $q$. The sliding ball analogy states that the front localizes where the two maximum values of the potential are equal. (c) Sketch of the bifurcation diagram and traveling wave solution of Eq. (\ref{eq:rde}). Blue lines denote stable solutions while the dashed (red) line corresponds to the unstable branch. Wave fronts (black lines and shaded area) penetrating the bistable region slow down and eventually come to rest at a stable fixed point of the front dynamics.  (d) Phase diagrams of possible parameter values allowing wave localization. The parameter range of wave localization increases with the Hill coefficient $n$.}
\end{center}
\end{figure*}

Specifically, we consider a one-dimensional system where diffusing particles are subject to three types of reactions: First, there are gain processes with a concentration-dependent rate that accounts for self-activation in gene regulatory systems or reproduction in population dynamics. Typically, these rates are small for low concentrations, then rise and finally saturate at high concentrations. In populations dynamics, this is referred to as the strong Allee effect \cite{Stephens:1999,ELE:ELE787}. In gene regulation, it can be due to cooperative transcription factor binding to a gene promoter. A common choice for the overall reaction rate is $k_r R_{a_0}^n(a)$ with the Hill function $R_{a_0}^n(a)\equiv\frac{a^n}{a_0^n+a^n}$, $k_r$ the maximum intrinsic production rate, and $a$ the particle concentration. The Hill coefficient $n$ measures the degree of cooperative binding in the promoter region, or, in ecology, the strength of an Allee effect. Second, we account for loss processes, where particles vanish with a certain rate $\lambda$. Third, in addition to self-activation, there may also be external sources for particle production. Here, we are interested in systems where this source is position-dependent and characterized by the overall rate $k_M M(x)$. The prefactor $k_M$ denotes the maximum rate of external activation, and $M(x)$ is a monotonically decreasing positive density profile with normalization $M(0)=1$. In the simplest case, where the profile results from a source-degradation dynamics~\cite{Grimm2010,Wartlick2011}, it is exponential $M(x)=e^{-x/\xi}$ with the decay length $\xi$, cf.  Fig.~\ref{fig:box1}~(a). Prominent examples are the concentration profile of Bicoid in \emph{Drosophila}~\cite{Grimm2010} and temperature or nutrient gradients in population dynamics~\cite{Tobin2007}. Since the production of Hunchback by Bicoid is mediated by cooperative binding, the profile $M(x)$ entering the overall production rate is commonly described by $M(x) \sim R^m_{I_0}(e^{- x/\xi})$~\cite{Burz1998}. The exponentially decaying signal induced by maternal source-degradation dynamics serves as an input to the gene regulation system. The latter is described by a Hill coefficient $m$ typically in the range from $1$ to $5$, and an activation threshold $I_0$.

In the limit of a large system size, fluctuations are of minor importance and the spatio-temporal dynamics is then aptly described by a reaction-diffusion equation, which in dimensionless form reads
\begin{equation}
\partial_t u =f(u,x)+\partial_{x x} u \, . \label{eq:rde}
\end{equation}
Here $f(u,x) \equiv r R^n_{u_0} (u) + M(x) - u$ comprises self-activation, external activation and degradation.  Concentration $u$, time $t$, and space $x$ are measured in units of  $k_M / \lambda$, $1/\lambda$ and $\sqrt{D/\lambda}$, respectively. The ratio $r\equiv k_r/k_M$ denotes the relative amplitude of self-activation and external activation mediated through $M(x)$.

\begin{figure*}[t]
\begin{center}
\includegraphics[width=0.99\textwidth]{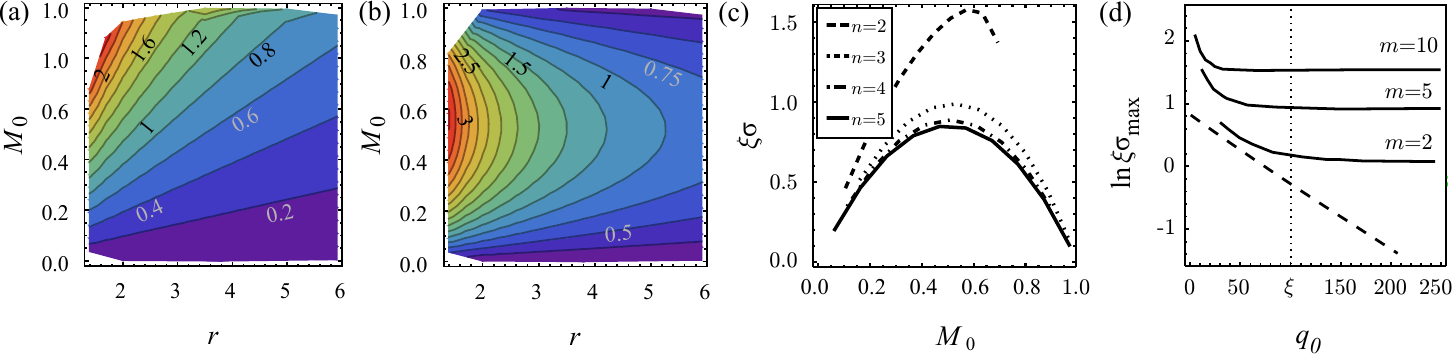}
\caption{(Color online)\label{fig:box2}  
Stability $\xi\sigma$, normalized to the steepness of the external profile, for (a) exponential and (b) sigmoidal [$m=5$] external profiles; the Hill coefficient for self-activation is $n=5$. Stability increases from blue to red: values of $\xi\sigma$ on lines of equal stability are indicated in the graph. While in both cases stability is optimized for weak self-activation $r$, they differ in the spatial position of the localized front as measured by the value of $M_0$.  (c) For sigmoidal profiles, small Hill coefficients $n$ for self-activation are optimal for front stability.  Parameters for plots (a-c) were $\xi = 100$, and $k=0.2$. (d) To study if regulation of an exponential  signal is biologically beneficial we determined the optimal stabilities which can be achieved for a front localized at a specific position. For each $q_0$ there are parameters $r$, $k$ and $M_0$ such that the linear stability $\sigma$ is maximal. Parameters were $n=5$, $2\leq r\leq 6$, $0.1\leq k\leq 1$. The plot shows the corresponding optimal values for $\sigma$ for exponential (dashed line) and sigmoidal external profiles (solid line, $m$ indicated in the graph). Sigmoidal gradients are generally more stable and in addition allow stable localization of fronts in a significant distance from the gradients source at $x=0$.
}
\end{center}
\end{figure*}
Traveling wave solutions of Eq.~\eqref{eq:rde} may be localized due to the combined effect of spatially varying external sources and bistability~\cite{Keitt2001,Mori2008,Mori2010}.  The basic mechanisms can be best understood in terms of the well-known sliding ball analogy \cite{Cross2009}, which here is complicated by the fact that the reaction term is space-dependent. Since in most biological situations a steep profile in $u$ is induced by a smooth external profile $M(x)$, we may assume a separation of length scales $\xi \gg \sqrt{D / \lambda}$ and $\xi$ much smaller than the system size. Then one can make a generalized traveling wave ansatz $U = U(x-q(t), y)$, where $x$ is a fast varying variable describing changes in the concentration profile, $y=x/\xi$ is a slowly varying variable describing changes in the external profile $M(x)$, and $q(t)$ denotes the front position. To leading order this gives  
\begin{equation}
- \dot q \partial_{x} U = \partial_{xx} U + \partial_U V(U,y) + {\mathcal O}({\xi^{-1}}) \, ,
\label{eq:rollball}
\end{equation}
which may be interpreted as a force balance for a particle (sliding ball) with mass $1$, friction $\dot q$ and potential $V(u,y)=\int^u f(\tilde u,y) \upd \tilde u$. Importantly, the potential parametrically depends on $y$, see Fig.~\ref{fig:box1}(b). For parameter regimes where $V$ has two maxima at $u^+(x)$ and $u^-(x)$, and a local minimum at $u^s(x)$, the velocity $\dot q$ must be chosen such that the sliding ball starting from the upper branch $u^+$ ends up at the lower branch $u^-$. The front speed is proportional to the difference between the two maxima of $V(u,y)$ and becomes zero if the condition $\Delta V(y)\equiv \int_{u^-}^{u^+} f(u,y)\upd u=0$ is satisfied. More quantitatively, following standard steps \cite{Cross2009,AKAbramyan2008,Vakulenko2009},  one finds~\footnote{One may also arrive at this equation employing a variational ansatz (Whitham principle)~\cite{AKAbramyan2008,Vakulenko2009}.}
\begin{equation}
\dot{q}\approx \frac{\Delta V(q)}{\int_{-\infty}^\infty [\partial_x U(x-q,y)]^2\upd x}\equiv c(q) \, ,
\label{eq:vel}
\end{equation}
where $U$ is the traveling wave solution.
The denominator roughly equals the maximum steepness of the front profile, and implies that steep fronts move slower~\cite{Cross2009}. 

In our class of models, a single branch of stable solutions at high concentrations typically undergoes a fold bifurcation for growing $x$, where the system is bistable on a confined spatial interval, see Fig.~\ref{fig:box1}~(c). For large $x$, a single branch at low concentrations remains. Within the bistable regime, the velocity $c(q)$ may change sign and thereby lead to a localization of the traveling wave front. 

We first determine the localization position $q_0$ of the front from $\Delta V(q_0)=0$. Approximations for $u^\pm (x)$ can be obtained by expanding  $f$ as Taylor or Laurant series: $u^-(x) = M(x) + \mathcal{O}(u^n)$ and $u^+(x) = M(x) + r +\mathcal{O}(u^{-n})$. For a given external profile $M(x)$, the potential reads $V(u,x) = -u\left[\slfrac{u}{2} -M(x)-r+r F\left(\slfrac{u^n}{u_0^n}\right)\right]$, where $F(z)\equiv {}_2F_1(1, \slfrac{1}{n}, 1+\slfrac{1}{n},-z)$ and ${}_2 F_1$ signifies the hypergeometric function. Keeping the dominant terms of $M(x)$ in $\Delta V$ we then obtain an expression for $M_0\equiv M(q_0)$ determining the localization position $q_0$:
\begin{equation*}
M_0\approx\frac{1}{2} r \left[1+\left(\frac{r}{u_0}\right)^n\right] \left(\frac{u_0}{r}\right)^n \left[2 F\fbrac{\frac{r^n}{u_0^n}}-1\right]\, . 
\end{equation*}
This is well approximated by a linear function of the form $g(n)\cdot(u_0-r/2)$ and converges to $u_0-\frac{1}{2}r$ for $n\rightarrow\infty$. For exponentially decaying gradients, the equilibrium front position is then given by $q_0=\xi\ln M_0$. For sigmoidal gradients, $M(x)\equiv \tilde{k} \cdot R^m_{k}(e^{- x/\xi})$ with dimensionless threshold $k$ and normalization factor $\tilde{k}\equiv k^m+1$, the front localizes at $q_0= \frac{\xi}{m} \ln\Bigl[\frac{ \tilde{k}-M_0}{\left(\tilde{k}-1\right)M_0}\Bigr]$.

Under which conditions is wave localization possible and robust? In \emph{Drosophila}, the parameters $r$, $u_0$ and $n$ are of special importance as they are main determinants of the gene regulation network \cite{Lopes2008}. The wave localizes if there is a bistable region in the bifurcation diagram,  \emph{i.e.}\/ if, for some $x$, the reaction term in Eq. (\ref{eq:rde})  has three real roots. Such values of $x$ exist if the maximum value of the derivative of $R_n^{u_0}(u) - u$ is greater than zero. We obtain as an approximate expression for the phase boundaries
\begin{equation}
-\left[F^{-1}\fbrac{1/2}\right] ^{\frac{1}{n}}\lessapprox \frac{u_0}{r} \lessapprox \frac{ n^2-1}{4 n}\left(\frac{n+1}{n-1}\right)^{1/n},
\end{equation}
and $\frac{u_0}{r}\lessapprox\frac{1}{2} +\frac{1}{r}$. Figure \ref{fig:box1}(d) shows that the range of allowed parameters grows with $n$. For large  $n$, the phase boundaries are well approximated by $\frac{1}{2}\leq\frac{ u_0}{r} \leq \frac{1}{2}+\frac{1}{r}$.  This translates to $a_0\lambda\approx k_r$, \emph{i.e.}\/  for front localization the overall degradation rate at the threshold must be of the same order as the maximum production rate.

To be stable against extrinsic perturbations the front should both relax back quickly into its equilibrium position and be insensitive to perturbations in the driving signal $M(x)$. Since a high relaxation rate implies that a front can follow changes in the signal quickly, the two stability criteria seem to be somewhat at odds. However, as shown below, they are in full accordance with the latter being less restrictive. 

The relaxation rate of the front back into its equilibrium position $q_0$ can be assessed within the framework of a linear stability analysis. Mathematically, this is given by expanding Eq.~\eqref{eq:vel} at $q_0$: $c(q)=-\sigma (q-q_0)+\mathcal{O}(q-q_0)^2$, where $\sigma\equiv -\left.\partial_q c(q)\right|_{q=q_0}$. The quantity $\sigma$ measures the stability of the fixed point $q_0$, such that large values of $\sigma$ correspond  to a stably localized front. We find
\begin{eqnarray}
\sigma =\left. -\frac{\partial_{M\fbrac{q}}\Delta V\fbrac{M(q)}\, \cdot \partial_q M(q)}{\int_{-\infty}^\infty [\partial_x U(x-q)]^2\upd x}\right|_{q=q_0} \,  \label{eq:sigma}
\end{eqnarray}
revealing that extrinsic stability is determined by three factors: In the numerator, the first factor describes how sensitively the potential difference of the stable states depends on the external source. The second factor, $\mu \equiv \left|\slfrac{\partial M(q) }{\partial q}\right|_{q_0}$, gives the steepness of the external profile at the localization position. While, therefore, a steeper source profile enhances front stability, the steepness of the front profile (denominator) has the opposite effect. The reason simply is that steeper fronts move slower and therefore also relax back more slowly, cf. Eq.~\eqref{eq:vel}. 

Figure \ref{fig:box2} shows the results of the numerical evaluation of $\sigma$ for both types of external sources; analytical results are given in the Supplemental Material. For both types of gradients we find that the localized wave front is most stable if $r$ is small, \emph{i.e}\/ if self-activation is weak or birth rates are low compared to the strength of the external source [Figs.~\ref{fig:box2}~(a) and (b)]. This can mainly be attributed to a decreased front steepness: reducing self-activation relative to external activation decreases the distance between the fixed points $u^\pm$ and thereby the steepness of the wave front. The front's stability is further optimized if it is localized at the steepest position of the external signal. For signals with a sigmoidal profile, this corresponds to $M_0 \approx 1/2$, and with $M_0\approx u_0-r/2$ in dimensionless form, it implies a relation between the degradation rate and the activation rates: $a_0\lambda = (k_r +k_M)/2$. Similarly, for an exponential profile with $M_0=1$ one finds  $a_0 \lambda = k_r/2 + k_M$.

How does cooperative binding influence stability? Since cooperativity in the kinetics of the external source implies a steeper sigmoidal profile, large values for the Hill coefficient $m$ increase the front's stability; see also the explicit expression for $\sigma$ in the Supplemental Material. Conversely, we find that stability is optimized for small values of $n$, \emph{i.e}\/ a low degree of cooperativity in the self-activation reaction [Fig. \ref{fig:box2}~(c)]~\footnote{This statement applies for exponential as well as sigmoidal external profiles}. This somewhat counterintuitive result can be attributed to a less steep front profile for small $n$; see Supplemental Material. Experimental data for Hunchback indeed indicate that the Hill coefficient $n$ for self-activation is rather low~\cite{Treisman1989, Lopes2008}. Figure~\ref{fig:box2}~(d) shows that stability for sigmoidal external gradients is, all other things being equal, generally higher than for exponential gradients. This implies that regulating an external positional signal is advantageous to the front's stability, since in this case the non-linear amplification of the signal makes it possible to create a steep signal even far away from the origin. 

To ensure stable localization, the front must also be robust against perturbations in $M(x)$. Specifically, it's position $q_0$ should only weakly depend on the local signal strength:  $\left| \slfrac{\partial q_0 (M) }{\partial M}\right|_{M_0} \ll 1$. This condition is equivalent to a steep source profile, $\mu = \left|\slfrac{\partial M(q) }{\partial q}\right|_{q_0} \gg 1$, and hence in full accordance with a large relaxation rate $\sigma$. It is, however, less restrictive since it is indifferent to changes in parameters that mainly affect the shape of the front, \emph{e.g.} the rate of self-activation $r$ and the Hill coefficient $n$; see Supplemental Material.

In many applications the front serves as a signal for further downstream processes, \emph{e.g.} to determine stripe-like patterning of the \emph{Drosophila} embryo~\cite{Gregor2007,Surkova2008}. In those instances it is also important that a front is not only stable against perturbations, but also sharply distinguishes between active and inactive regions. This requires a steep front which is generally obtained if self-activation is strong compared to external activation and, to a lesser degree, if binding cooperativity is strong; see Supplemental Material. Sharp fronts, however, are susceptible to extrinsic fluctuations and one has to sacrifice front stability for the precision of the transmitted signal.

Intrinsic noise resulting from small copy number fluctuations also affects the stability of the localized wave front. In this case, stability can be measured in terms of the  ratio $D/D_f$ between the individual particle's and the front's diffusion constants. The latter can be calculated following the steps outlined in Ref. \cite{Meerson2011a}: 
\begin{equation}
\frac{D}{D_f} = \biggl.\frac{N \cdot \left[ \int_{-\infty}^{\infty}\upd x (U')^2\right]^2}{\int_{-\infty}^{\infty} \upd x \left[ \frac{1}{2}(U')^2 h(U)+ U  (U'')^2\right]}\biggr|_{q=q_0} \, , \label{eq:wkb}
\end{equation}
where $h(U) \equiv  R^n_{u_0} (U) + M(x) + U$, and $U$ denotes the stationary solution. Generally, the front's diffusion constant is smaller than the particle's diffusion constant by a factor $N$, which corresponds to the typical number of particles in the front region. The integral in the numerator gives the maximum steepness of the front. Hence, as opposed to extrinsic stability, intrinsic stability is optimal for steep fronts. Shallow fronts are prone to stochastic switching, as the entropy barrier between the stable states is reduced in the front region. The terms in the denominator account for the reaction and diffusion noise. In contrast to extrinsic stability, we here find that the front is most robust against fluctuations for strong self-activation $r$. The reason for this is that, as $r$ determines the amount of reactions necessary to locally switch between the stable states, the rate of stochastic switching decreases for large $r$. Explicit analytical results can be found in the Supplemental Material. 

In conclusion, we identified conditions optimizing the stability and robustness of localized wave fronts for different types of perturbations. We find that increasing cooperativity in self-activation broadens the parameter regime where wave localization becomes possible and thereby increases the robustness of the localization mechanism. Interestingly, there is a tradeoff between the stability of the wave front to extrinsic and intrinsic perturbations. While weak self-activation or low birth rates enhance the stability with respect to extrinsic perturbations, stochastic defocussing is minimized for strong self-activation. The latter also increases the spatial precision of the signal transmitted by the front to downstream processes. Moreover, we showed that processing input from external sources with a cooperative gene activation mechanism generally enhances the front's stability even far away from the source. Surprisingly, while cooperativity in external activation increases the front's stability with respect to extrinsic perturbations, the opposite holds true for self-activation.

The conflict between intrinsic and extrinsic stability affects, for example, the design of gene circuits in developmental systems. Our results suggest different design principles depending on the particle number. If the number of involved particles is large, intrinsic noise is irrelevant. Then the parameters of the genetic network may be optimized for robustness against external perturbations which is achieved by weak self-activation and strong cooperativity in external activation. Conversely, if particle numbers are low, robustness against intrinsic noise requires strong and cooperative self-activation. To also safeguard against external perturbation then requires additional mechanisms beyond those included in our simplified model. We expect these general results to be important guiding principles in the context of biological pattern forming systems, such as cell polarization or the segmentation of embryos.

\begin{acknowledgments}
This research was supported by the German Excellence Initiative via the program `Nanosystems Initiative Munich' and the German Research Foundation via the SFB 1032 ``Nanoagents for Spatiotemporal Control of Molecular and Cellular Reactions''. 
\end{acknowledgments}

%

\end{document}